\documentclass{article}

    \PassOptionsToPackage{numbers, compress}{natbib}

\usepackage[preprint]{neurips_2025}

\usepackage[utf8]{inputenc}
\usepackage[T1]{fontenc}
\usepackage{hyperref}
\usepackage{url}
\usepackage{booktabs}
\usepackage{amsfonts}
\usepackage{amsmath}
\usepackage{nicefrac}
\usepackage{microtype}
\usepackage{multirow}
\usepackage{xcolor}
\usepackage[pdftex]{graphicx}
\usepackage{enumitem}
\usepackage{array}
\usepackage{pifont}
\usepackage{todonotes}
\usepackage{longtable}
\usepackage{tabularx}
\usepackage{ragged2e}
\usepackage[most]{tcolorbox}

\title{QuantMind: A Context-Engineering Based Knowledge Framework for Quantitative Finance}

\author{Haoxue Wang\textsuperscript{\rm 1, \rm 2}\thanks{Equal Contribution.}\hspace{4pt},
Keli Wen\textsuperscript{\rm 1}\footnotemark[1]\hspace{4pt},
Yuante Li\textsuperscript{\rm 1, \rm 3}\footnotemark[1]\hspace{4pt},
Qu Qianchen \textsuperscript{\rm 1, \rm 4}\footnotemark[1]\hspace{4pt},
Xiangxu Mu \textsuperscript{\rm 1, \rm 5}\footnotemark[1]
\\
\textbf{Xinjie Shen \textsuperscript{\rm 1}, Jiaqi Gao \textsuperscript{\rm 1}, Chenyang Chang \textsuperscript{\rm 1}, Chuhan Xie \textsuperscript{\rm 1, \rm 6}, San Yu Cheung \textsuperscript{\rm 1, \rm 7}} \\
\textbf{Zhuoyuan Hu \textsuperscript{\rm 1, \rm 8}, Xinyu Wang\textsuperscript{\rm 1, \rm 9}, Sirui Bi \textsuperscript{\rm 1, \rm 2}, Bi'an Du \textsuperscript{\rm 1}}\thanks{Corresponding Author.} \\
\textsuperscript{\rm 1} LLMQuant Research, 
\textsuperscript{\rm 2} University of Cambridge,
\textsuperscript{\rm 3} Carnegie Mellon University \\
\textsuperscript{\rm 4} National University of Singapore,
\textsuperscript{\rm 5} IEIT SYSTEMS CO., LTD.,
\textsuperscript{\rm 6} Peking University \\
\textsuperscript{\rm 7} The Chinese University of Hong Kong,
\textsuperscript{\rm 8} Shanghai Jiao Tong University,
\textsuperscript{\rm 9} Beihang University\\
\texttt{\{haoxue, keli\}@llmquant.com}, \texttt{yuantel@cs.cmu.edu}\\
\texttt{e1486355@u.nus.edu},
\texttt{muxiangxu@ieisystem.com}, \texttt{bian@llmquant.com}
}

\begin{document}
\maketitle

\begin{abstract}
Quantitative research increasingly relies on unstructured financial content such as filings, earnings calls, and research notes, yet existing LLM and RAG pipelines struggle with point-in-time correctness, evidence attribution, and integration into research workflows.
To tackle this, We present \texttt{QuantMind}, an intelligent knowledge extraction and retrieval framework tailored to quantitative finance.
\texttt{QuantMind} adopts a two-stage architecture: (i) a \textbf{knowledge extraction} stage that transforms heterogeneous documents into structured knowledge through multi-modal parsing of text, tables, and formulas, adaptive summarization for scalability, and domain-specific tagging for fine-grained indexing; and (ii) an \textbf{intelligent retrieval} stage that integrates semantic search with flexible strategies, multi-hop reasoning across sources, and knowledge-aware generation for auditable outputs.
A controlled user study demonstrates that \texttt{QuantMind} improves both factual accuracy and user experience compared to unaided reading and generic AI assistance, underscoring the value of structured, domain-specific context engineering for finance.
\end{abstract}

\vspace{-0.25em}
\section{Introduction}

The growing reliance on unstructured financial data, including SEC filings, earnings call transcripts, and broker research notes, has fundamentally influenced quantitative finance. Domain-specific language models like FinBERT~\citep{araci2019finbert}, BloombergGPT~\citep{wu2023bloomberggpt}, and FinGPT~\citep{yang2023fingpt}, together with retrieval-augmented generation (RAG) pipelines~\citep{lewis2020retrieval,karpukhin2020dpr,khattab2020colbert}, have advanced financial NLP and retrieval. However, they continue to face persistent challenges: \ding{182} lack of point-in-time correctness, \ding{183} insufficient evidence traceability, and \ding{184} limited integration with quantitative research workflows. These limitations restrict their reliability in practical financial applications.

The rise of long-context and agent-oriented large language models (LLMs), such as GPT-4~\citep{Achiam2023GPT4TR} and Claude~\citep{claude3}, highlights the need for structured, agent-readable knowledge frameworks in finance. While general frameworks such as \texttt{agents.md}~\citep{openai_2025} and \texttt{claude.md}~\citep{claude_code_best_practices_2025} provide broad design principles, they do not address the domain-specific requirements of financial research. Meeting these demands requires transforming heterogeneous and dynamic financial artifacts into structured, auditable knowledge that supports reproducible reasoning and seamlessly integrates with customizable research workflows.

To address these gaps, we propose \texttt{\textbf{QuantMind}}, an intelligent knowledge extraction and retrieval framework tailored to quantitative finance. Our contributions are threefold:
\vspace{-0.12cm}
\begin{itemize}[leftmargin=4ex]
    \item We design a two-stage decoupled architecture comprising \textbf{knowledge extraction} and \textbf{intelligent retrieval}, enabling point-in-time correctness, provenance preservation, and reproducibility.  \vspace{-0.06cm}
    \item In the extraction stage, we introduce AI-driven multi-modal parsing, adaptive summarization with cost-optimized chunking, and domain-specialized tagging for fine-grained indexing. \vspace{-0.06cm}
    \item In the retrieval stage, we develop a flexible retrieval pattern architecture supporting both DeepResearch and RAG, unified knowledge representation with vectorization readiness, and a flow-based orchestration layer inspired by multi-agent systems.
\end{itemize}

\section{Related Work}

\textbf{Retrieval-Augmented Generation and Context Engineering.}
RAG combines a retriever and a generator, typically built on dense encoders (e.g., SBERT~\citep{reimers2019sentence}) and ANN indexes (e.g., FAISS~\citep{douze2025faisslibrary}). Standard pipelines operate in a straightforward manner: documents are chunked, embedded, and the top-$k$ results are concatenated~\citep{chen-etal-2017-reading,lewis2020retrieval}. This design often produces weak provenance. To address these limitations, context engineering introduces reranking~\citep{mei2025surveycontextengineeringlarge,nogueira2020passagererankingbert}, multi-hop retrieval~\citep{santhanam-etal-2022-colbertv2,yang-etal-2018-hotpotqa}, and structure-aware methods~\citep{qi-etal-2019-answering,ijcai2022p629}. Nevertheless, existing systems remain fragment-centric and version-insensitive. In financial applications, where critical evidence is distributed across tables and figure captions, such fragmentation undermines reference integrity and disrupts contextual coherence.

\textbf{Scientific Knowledge Management Systems.}
Beyond finance, infrastructures such as the Semantic Scholar Literature Graph, S2ORC~\citep{lo2020s2orc}, and ORKG~\citep{jaradeh2019orkg} demonstrate how structured corpora and knowledge graphs can facilitate search and enable limited forms of reasoning through explicit structure and versioning~\citep{priem2022openalex}. However, these systems are domain-general and do not provide an agentic, finance-aware layer capable of multistep, auditable, retrieval-augmented reasoning over continuously evolving documents.

\vspace{-0.5em}

\section{QuantMind}
\label{sec:system}

\begin{figure}[!h]
  \centering  \includegraphics[width=0.8\linewidth]{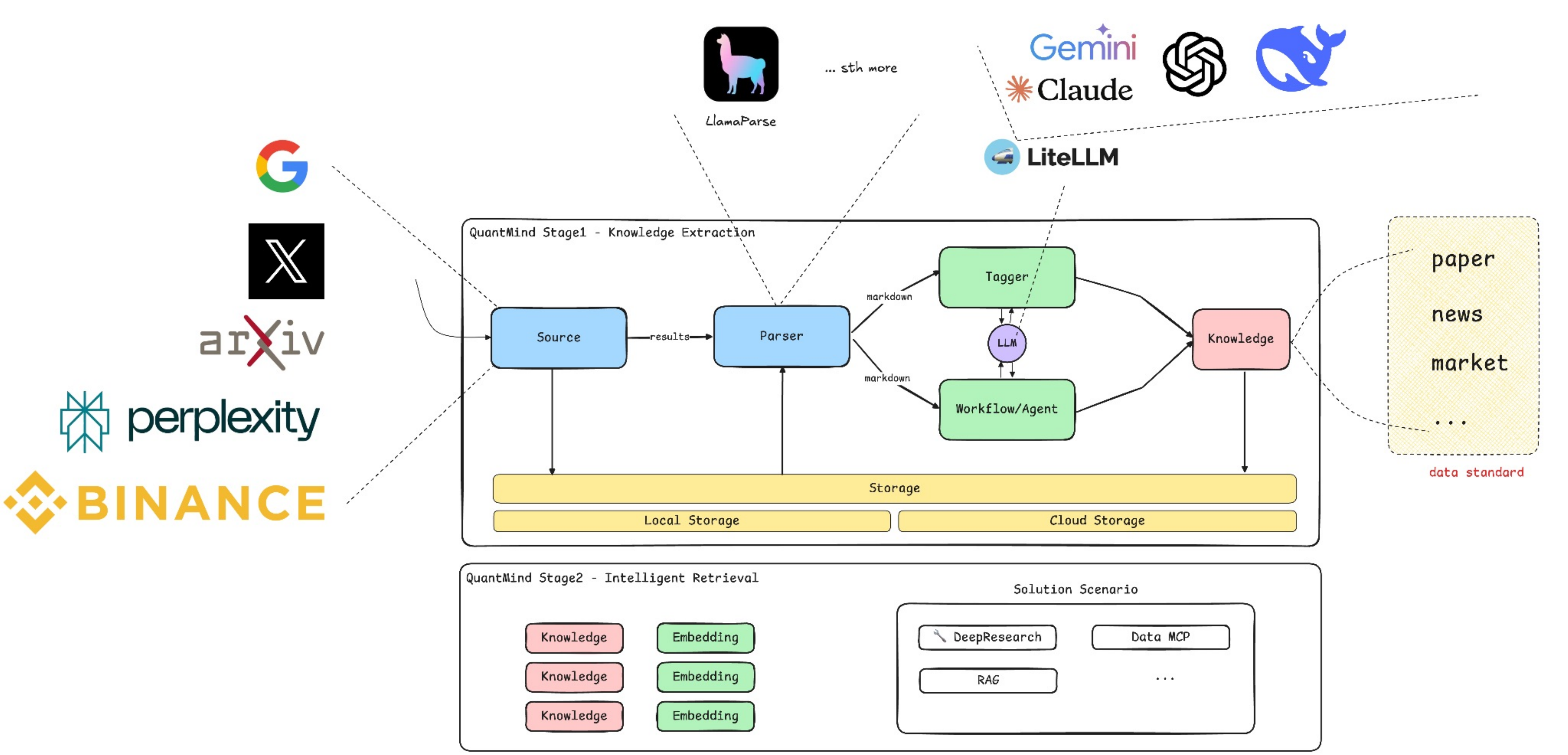}
  \caption{Architecture of \texttt{QuantMind}. The framework follows a two-stage design: (i) \textbf{knowledge extraction}, which structures heterogeneous financial documents via multi-modal parsing, summarization, and tagging; and (ii) \textbf{intelligent retrieval}, which supports semantic search, multi-hop reasoning, and knowledge-aware generation for quantitative research.}
  \label{fig:arch}
\end{figure}

QuantMind adopts a two-stage decoupled architecture that separates knowledge extraction from intelligent retrieval, as illustrated in Figure~\ref{fig:arch}. The framework consists of two main stages: 1) Knowledge Extraction Stage, which transforms unstructured financial content into structured knowledge units, and 2) Intelligent Retrieval Stage, which enables dynamic retrieval patterns for knowledge access and analysis.

\vspace{-0.12cm}
\textbf{Stage1: Knowledge Extraction.} Reliable LLM analysis, reasoning, and QA depend on the quality of source data \citep{zhou2025surveyllmtimesdata, zhang2024document}. We therefore build a knowledge extraction pipeline that starts with large scale crawling and continues with multimodal parsing and enrichment: 

\ding{223} \textit{Multi-Modal Parsing.} In quantitative finance, key insights are often encoded not only in text but also in tables of results, figures of market dynamics, formulas defining models, and even podcast and video modalities, ignoring these modalities risks of incomplete or biased knowledge extraction. To address this, we adopt a multi-modal parsing operator $\mathcal{P}$ (implemented with LlamaParse that decomposes a financial document $D$ into $\mathcal{P}(D)=\{T,F,M,S\}$, where $T$ denotes textual passages, $F$ visual and tabular elements, $M$ symbolic mathematical content, and $S$ the semantic organization that integrates them.
By preserving cross-modal dependencies, the extracted knowledge serves as an abstraction that encodes both the content and the underlying logical–empirical relations of the paper.

\ding{223} \textit{Adaptive Summarization.} 
Long documents are costly to be processed directly.
To improve scalability, we split the content $C=\{c_1,\ldots,c_n\}$ into semantically coherent segments and summarize each segment with a lightweight model $s_i=M_{\text{cheap}}(c_i)$. 
These local summaries are then aggregated by a more expressive model $S_{\text{final}}=M_{\text{powerful}}(s_1\oplus\cdots\oplus s_n)$, where $\oplus$ denotes ordered concatenation.
This two-stage design reduces cost since $\text{Cost}_{\text{total}}=n \cdot \text{Cost}(M_{\text{cheap}})+\text{Cost}(M_{\text{powerful}})\ll n\cdot \text{Cost}(M_{\text{powerful}})$, while retaining fidelity at the global level.

\ding{223} \textit{Domain-Specialized Tagging.} 
To better capture relationships across papers, we introduce a high-dimensional tagging scheme that provides fine-grained indexing.
Each paper $P$ is enriched with structured labels covering its primary research area, secondary topics, methodological orientation, and application domain.
For robustness, each tag $t_i$ is associated with a confidence score $c_i=f(C,t_i,M_{\text{tag}})\in[0,1]$, computed by a discriminative tagging model $M_{\text{tag}}$, which supports domain-aware retrieval and systematic cross-paper comparison.

\textbf{Stage2: Intelligent Retrieval.}
Once high-quality knowledge units are extracted, the next challenge lies in retrieving and integrating them for downstream reasoning and question answering.
This stage equips the framework with flexible retrieval patterns, multi-hop reasoning, and knowledge-aware generation.

\ding{223} \textit{Adaptive Retrieval Strategies.} 
Given a query $q$ and a knowledge base $K$, the framework dynamically selects retrieval patterns. 
For simple factual queries, a lightweight RAG-style approach retrieves once, $R=\mathrm{Retrieve}(q,K)$, and augments generation with this context to produce an answer.  
For more complex reasoning, a DeepResearch process iteratively expands context,  
$
R_t = \mathrm{Retrieve}(q \oplus R_{t-1}, K), \quad R_{\text{final}}=M_{\text{reason}}(q,R_1,\ldots,R_n),
$
balancing efficiency with the need for depth.  

\ding{223} \textit{Multi-Hop Reasoning.} 
Certain queries require connections that span multiple documents or methodologies. 
To support this, the system performs iterative retrieval and synthesis across hops, progressively enriching the query until it captures the necessary context. 
This mechanism enables conceptual linking (e.g., relating market regimes), methodological comparison (contrasting algorithms), and empirical validation across independent studies.

\ding{223} \textit{Knowledge-Aware Generation.} 
The retrieved context $R_{\text{context}}$ is integrated with the query to enhance both the generation of questions and answers. 
Enriched questions $Q_{\text{enhanced}}=M_{\text{gen}}(q\oplus R_{\text{context}})$ incorporate broader connections, while answers $\text{Ans}=M_{\text{ans}}(Q_{\text{enhanced}},q\oplus R_{\text{context}})$ combine local content with the retrieved knowledge.  
This allows the system to move beyond factual responses, supporting comparative reasoning, cross-domain insights, and temporally informed analysis. 

\vspace{-0.25em}
\section{User Study}
\label{sec:study}

\textbf{Goal and Design.} The aim of this study is to assess whether \texttt{QuantMind} improves quantitative finance research performance relative to unaided reading and a generic AI assistant.
We define research performance as the ability to:  
(i) extract factual information from academic papers (e.g., factor definitions, alpha sources, return characteristics), and  
(ii) perform higher-level reasoning (e.g., evaluating factor generalization across markets, horizons, or methodological extensions).

We employed a within-subjects repeated measures design with counterbalancing to control for potential learning and fatigue effects.
Each participant completed six tasks, each derived from a distinct finance paper, under one of the following three conditions:
\ding{182} \textbf{Without AI}: Participants relied exclusively on their own reading and research skills;
\ding{183} \textbf{With AI Assistant}: Participants were provided with assistance generated by a generic LLM (\texttt{GPT-4o}), using a fixed prompt created prior to the study to supply supplementary information without offering direct answers;
\ding{184} \textbf{With QuantMind}: Participants received assistance generated by the proposed \texttt{QuantMind} framework, which employed domain-specific RAG from a structured knowledge base. The prompting strategy was optimized to extract and synthesize relevant information from both the focal paper and related research.

To balance exposure across conditions, we applied a Latin Square design \citep{fisher1949design} with participants and papers as blocking factors, ensuring equal representation of each condition and controlling for order effects. The detailed assignment is provided in Appendix~\ref{Counterbalanced_Assign}.

\textbf{Corpus.} We curated a set of six seminal papers in quantitative finance (see Appendix~\ref{SRPQD}), selected to capture both historical breadth and methodological diversity. The corpus spans foundational contributions in financial economics, canonical studies in factor modeling, and recent works situated at the intersection of machine learning and quantitative finance. For each paper, we designed two distinct categories of evaluation tasks:  
\ding{182} \textbf{Information-extraction questions} (3–4 per paper): objective queries with verifiable ground-truth answers, such as precise definitions of factors, data sources, or performance statistics. Responses were scored against a pre-defined answer key.
\ding{183} \textbf{Analytical questions} (1–2 per paper): open-ended tasks requiring higher-order reasoning, including the evaluation of factor generalizability across markets, time horizons, or methodological extensions. Responses were assessed using an LLM-as-a-judge framework \citep{NEURIPS2023_91f18a12}, which enabled systematic comparison of logical coherence, depth, and interpretive accuracy.

\textbf{Metrics.} We evaluated research performance along two principal dimensions:
\texttt{Quality}: quality of answers judged upon subjects' comprehension, logic, as well as the breadth and depth of their insights.
\texttt{UX Rating}: subjective evaluation of AI assistance on a 5-point Likert scale covering 4 dimensions: relevance, accuracy, helpfulness, and clarity.

\textbf{Statistical Analysis.} We analyzed both performance metrics (\texttt{Quality} and \texttt{UX Rating}) using linear mixed-effects models to account for the within-subjects repeated measures design:  
\begin{equation}
Y_{ijk} = \mu + \tau_i + s_j + p_k + \epsilon_{ijk} \ ,
\end{equation}
where $Y_{ijk}$ is the observed value for treatment $i$, subject $j$, and paper $k$; $\mu$ is the grand mean; $\tau_i$ denotes the fixed effect of treatment condition; $s_j \sim N(0, \sigma_s^2)$ and $p_k \sim N(0, \sigma_p^2)$ are random intercepts for subject and paper, respectively; and $\epsilon_{ijk} \sim N(0, \sigma_\epsilon^2)$ is the residual error.

Such model specification allows us to isolate the treatment effect while accounting for subject- and paper-level variability.
To further assess differences between assistance conditions, we conducted post-hoc pairwise comparisons using Tukey’s HSD ($\alpha = .05$), and quantified effect sizes with Cohen’s $d$.

\textbf{Results.} Table~\ref{table-2} reports descriptive statistics and pairwise differences. Each treatment condition includes approximately equal numbers of observations (66–68), and the variance structure indicates heterogeneity across both participants and papers.  
For \texttt{Quality}, \texttt{QuantMind} significantly outperforms both baselines: a 1.14-point gain over the no-AI condition ($p < 0.001$) and a 0.43-point gain over the generic AI assistant ($p = 0.001$).
For \texttt{UX Rating}, while the overall treatment effect was not significant at the conventional level ($p = 0.108$), \texttt{QuantMind} improved average ratings by 0.38 points and yielded a statistically significant enhancement in perceived helpfulness ($p = 0.003$).
These findings suggest that \texttt{QuantMind} not only enhances the accuracy and depth of research outputs but also provides a more supportive user experience compared to both unaided reading and a generic AI assistant.
Complete statistical outputs, including confidence intervals and additional pairwise contrasts, are provided in Appendix~\ref{RER}.

\begin{table}[htb]
\centering
\label{table-2}
\resizebox{0.9\textwidth}{!}
{
\begin{tabular}{lcccccc}
\toprule
Metric & Treatment & $n$ & Mean & SD & Median & Min--Max \\
\midrule
\multirow{3}{*}{Accuracy (0--5)} 
& Without AI & 66 & 3.11 & 0.68 & 3.0 & 1.0--4.5 \\
& AI Assistant & 68 & 3.82 & 0.64 & 4.0 & 2.0--5.0 \\
& \textit{QuantMind} & 66 & 4.25 & 0.55 & 4.5 & 3.0--5.0 \\
\midrule
\multirow{3}{*}{UX Rating (1--5)} 
& Without AI & 66 & --- & --- & --- & --- \\
& AI Assistant & 68 & 3.78 & 0.62 & 4.0 & 2.0--5.0 \\
& \textit{QuantMind} & 68 & 4.21 & 0.55 & 4.0 & 3.0--5.0 \\
\bottomrule
\end{tabular}
}
\vspace{0.25cm}
\caption{Descriptive statistics of accuracy and UX ratings across treatment conditions, showing $n$, mean, SD, median, and range.}
\end{table}

\vspace{-0.25em}
\section{Conclusion}
We introduced \texttt{QuantMind}, a framework that converts heterogeneous financial artifacts into a structured, agent-readable knowledge base with domain-aware retrieval. Its decoupled extraction and retrieval stages ensure point-in-time correctness, provenance, and reproducibility, while supporting DeepResearch, RAG, and structured natural language access. A controlled study confirms that \texttt{QuantMind} improves both accuracy and efficiency in quantitative research.

\bibliographystyle{unsrtnat}
\bibliography{neurips_2025.bib}

\newpage
\appendix
\section{Detailed Experimental Results}
\label{DER}
\subsection{Prompt Design}

We detail the prompts used in the experiment to ensure consistency across conditions. 
The \emph{AI Assistant Prompt} provides generic background information, the \emph{QuantMind Structured Prompt} adds domain-specific context from the knowledge base, and the \emph{LLM Judge Prompt} specifies the evaluation criteria for answer quality. 
Full prompt texts are shown below.

\begin{tcolorbox}[
  colback=gray!5,
  colframe=gray!80,
  title style={fontupper=\bfseries\large},
  title=AI Assistant Prompt,
  breakable
]
\label{tab:ai-prompt}
Based on the provided research paper, generate helpful background information and relevant concepts that could assist in answering the following question, but do not provide direct answers.
\end{tcolorbox}

\begin{tcolorbox}[
  colback=gray!5,
  colframe=gray!80,
  title style={fontupper=\bfseries\large},
  title=QuantMind Structured Prompt,
  breakable
]
\label{tab:quantmind-prompt}

Retrieve and synthesize relevant concepts, methodologies, and contextual information from the quantitative finance literature to provide comprehensive background support for addressing the following question, ensuring no direct answers are provided.
\end{tcolorbox}

\begin{tcolorbox}[
  colback=gray!5,
  colframe=gray!80,
  title style={fontupper=\bfseries\large},
  title=LLM Judge Prompt for Answer Quality Evaluation,
  breakable
]
\label{tab:llm-judge-prompt}

Please evaluate the quality of answers from the following 6 participants for the same academic question.\\
\\
Paper Context: {\{{paper\_context}\}} \\
Question: {\{{question\_text}\}} \\
Answers: {\{{formatted\_answers}\}}\\
\\
Please rate each participant's answer on a scale of $1$--$5$ (with one decimal point allowed) based on the following criteria:\\
\textbf{Scoring Criteria:}\\
$1$ point: Answer is completely irrelevant, incorrect, or incomprehensible\\
$2$ points: Answer is mostly irrelevant, showing little understanding of the question\\
$3$ points: Answer is partially relevant with some understanding but lacks depth\\
$4$ points: Answer is relevant with reasonable depth and clear logic\\
$5$ points: Answer is highly relevant, demonstrates deep analysis, rigorous logic, and unique insights\\
\textbf{Note:}\\
- For participants marked as ``No response,'' please return ``No response''\\
- Only rate participants who provided substantive answers\\
\\
Please return the scoring results in the following format:\\
Participant $1$: X.X points\\
Participant $2$: X.X points\\
Participant $3$: X.X points\\
Participant $4$: X.X points\\
Participant $5$: X.X points\\
Participant $6$: X.X points\\
\\
Please ensure that your evaluations are objective and impartial, based on the content quality, logical coherence, and relevance of each answer.
\end{tcolorbox}

\subsection{Counterbalanced Assignment in Within-subjects User Study}
\label{Counterbalanced_Assign}

To mitigate order and sequence effects, we adopted a counterbalanced assignment scheme based on an incomplete Latin square design. This approach ensured that each assistance condition (Without AI [W], Baseline AI Assistant [B], and QuantMind [C]) appeared with approximately equal frequency across participants and papers, while preventing systematic biases due to task ordering. Table~\ref{table:latin-square} illustrates the assignment matrix, where rows correspond to participants and columns to the six curated papers.  

\begin{table}[h!]
\centering
\begin{tabularx}{\textwidth}{lXXXXXX}
\toprule
Participant & Paper 1 & Paper 2 & Paper 3 & Paper 4 & Paper 5 & Paper 6 \\
\midrule
Subject 1 & W & W & B & B & C & C \\
Subject 2 & W & B & B & C & C & W \\
Subject 3 & B & B & C & C & W & W \\
Subject 4 & B & C & C & W & W & B \\
Subject 5 & C & C & W & W & B & B \\
Subject 6 & C & W & W & B & B & C \\
\bottomrule
\end{tabularx}
\vspace{0.15cm}
\caption{Counterbalanced assignment under an incomplete Latin square. Rows denote participants, columns denote papers, and entries indicate assistance conditions: Without AI (W), Baseline AI Assistant (B), or \texttt{QuantMind} (C).}
\label{table:latin-square}
\end{table}

\subsection{Selected Research Papers and Question Design}
\label{SRPQD}

Table \ref{tab:papers} presents the mapping between paper IDs and the corresponding evaluation questions used in the user study. 
For clarity, the six paper IDs correspond to the following works.
\textbf{P1}: Anomalies in the China A-share market \citep{JANSEN2021101607};
\textbf{P2}: Good volatility, bad volatility, and the cross section of stock returns \citep{Bollerslev_Li_Zhao_2020};  
\textbf{P3}: Overnight returns, daytime reversals, and future stock returns \citep{AKBAS2022850};
\textbf{P4}: Panel data nowcasting: The case of price–earnings ratios \citep{babii2023paneldatanowcastingcase};
\textbf{P5}: Replication of Reference-Dependent Preferences and the Risk-Return Trade-Off in the Chinese Market \citep{xu2025replicationreferencedependentpreferencesriskreturn}; and
\textbf{P6}: The pricing of options and corporate liabilities \citep{0b9b8115-a8b8-3422-8e1c-a62077de6621}.

\begin{longtable}{ p{1.0cm} p{1.5cm} p{9.5cm} }
\toprule
\textbf{Paper ID} & \textbf{Question ID} & \textbf{Question} \\
\midrule
\endfirsthead

\multicolumn{3}{c}%
{{\bfseries Table Continued}} \\
\toprule
\textbf{Paper ID} & \textbf{Question ID} & \textbf{Question} \\
\midrule
\endhead

\bottomrule
\endfoot

\bottomrule
\endlastfoot

1 & 1 & Based on the paper, which factors do you think perform strongest in the Chinese A-share market? How would you explain the source of these factor returns? \\
\midrule
1 & 2 & Which factors underperform in the paper? In your research or practice, how would you improve these factors or explore their potential value? \\
\midrule
1 & 3 & Apart from short-selling constraints, state-owned enterprises, and market reforms, what other China-specific factors might affect factor performance? \\
\midrule
1 & 4 & How do you assess the robustness and practicality of a factor in quantitative investment practice? \\
\midrule
1 & 5 & Does this paper provide clues about future research directions for factors? What new ideas do you have for future research? \\
\midrule
1 & 6 & If you were to construct a new factor from scratch, how would you use the methodology in this paper to validate its effectiveness? \\
\midrule
2 & 7 & What is the core idea of the paper? \\
\midrule
2 & 8 & What data is needed to construct trading signals? What are the time frequencies of these data? \\
\midrule
2 & 9 & Which stock market data was used for backtesting in the paper? What metrics were selected for the backtest? \\
\midrule
2 & 10 & Pseudocode for calculating the simplest version of the factor? \\
\midrule
2 & 11 & What is the source of returns for this factor? What type of factor is it? \\
\midrule
2 & 12 & Can other factors be derived from this factor? \\
\midrule
3 & 13 & What is the core idea of the paper? \\
\midrule
3 & 14 & What data is needed to construct trading signals? What are the time frequencies of these data? \\
\midrule
3 & 15 & Which stock market data was used for backtesting in the paper? What metrics were selected for the backtest? \\
\midrule
3 & 16 & Pseudocode for calculating the simplest version of the factor? \\
\midrule
3 & 17 & What is the source of returns for this factor? What type of factor is it? \\
\midrule
3 & 18 & Can other factors be derived from this factor? \\
\midrule
4 & 19 & What is the approach of the study for constructing predictive factors? \\
\midrule
4 & 20 & How is the sparse group LASSO (sg-LASSO) regularization method defined in the paper? Provide the formula and explain the parameters. \\
\midrule
4 & 21 & What are the inputs of the machine learning models used in the paper? \\
\midrule
4 & 22 & What are the limitations of the models in the paper? \\
\midrule
5 & 23 & What is the core idea of the paper? \\
\midrule
5 & 24 & What data is needed to construct trading signals? What are the time frequencies of these data? \\
\midrule
5 & 25 & Which stock market data was used for backtesting in the paper? What metrics were selected for the backtest? \\
\midrule
5 & 26 & Write pseudocode for calculating the simplest version of the factor. \\
\midrule
5 & 27 & What is the source of returns for this factor? What type of factor is it? \\
\midrule
5 & 28 & Can other factors be derived from this factor? \\
\midrule
6 & 29 & Describe the basic assumptions and derivation idea of the Black-Scholes option pricing model. \\
\midrule
6 & 30 & What parameters are set in the model? Please explain their economic meaning. \\
\midrule
6 & 31 & When the market exhibits a "volatility smile," how does the Black-Scholes model explain this deviation? \\
\midrule
6 & 32 & Provide the formula used for pricing with the Black-Scholes model. \\
\midrule
6 & 33 & What are the limitations of the Black-Scholes model in actual markets? \\
\midrule
6 & 34 & Assuming you are an options trader, how would you use the Black-Scholes model to identify overvalued or undervalued options? \\
\bottomrule
\\
\caption{Mapping of Questions to Paper IDs.}
\label{tab:papers}
\end{longtable}

\subsection{Raw Experimental Results}
\label{RER}
\subsubsection{Per-Participant Results}

Table~\ref{tab:ux_results} presents the UX rating result. The subject and paper ID columns in this table correspond to the respective columns in Table~\ref{table:latin-square}.

\begin{table}[h!]
\centering
\begin{tabular}{cccccccc}
\toprule
\textbf{subject} & \textbf{paper ID}  & \textbf{assistance lvl} & \textbf{relevance} & \textbf{accuracy} & \textbf{helpfulness} & \textbf{clarity} & \textbf{average} \\
\midrule
1 & 3 & 2 & 3 & 4 & 4 & 3 & 3.5  \\

1 & 4 & 2 & 4 & 4 & 4 & 3 & 3.75 \\

1 & 5 & 3 & 5 & 4 & 5 & 2 & 4 \\

1 & 6 & 3 & 5 & 4 & 5 & 3 & 4.25 \\

2 & 2 & 2 & 4 & 5 & 4 & 3 & 4 \\

2 & 3 & 2 & 5 & 4 & 4 & 4 & 4.25 \\

2 & 4 & 3 & 4 & 4 & 4 & 3 & 3.75 \\

2 & 5 & 3 & 4 & 4 & 5 & 4 & 4.25 \\

3 & 2 & 2 & 2 & 2 & 1 & 1 & 1.5 \\

3 & 1 & 2 & 4 & 4 & 2 & 2 & 3 \\

3 & 3 & 3 & 2 & 3 & 2 & 2 & 2.25 \\

3 & 4 & 3 & 4 & 3 & 4 & 3 & 3.5 \\

4 & 1 & 2 & 4 & 4 & 3 & 4 & 3.75 \\

4 & 6 & 2 & 5 & 4 & 4 & 5 & 4.5 \\

4 & 2 & 3 & 5 & 5 & 5 & 5 & 5 \\

4 & 3 & 3 & 5 & 5 & 5 & 5 & 5 \\

5 & 5 & 2 & 4 & 5 & 5 & 5 & 4.75 \\

5 & 6 & 2 & 4 & 3 & 4 & 3 & 3.5 \\

5 & 1 & 3 & 5 & 4 & 5 & 4 & 4.5 \\

5 & 2 & 3 & 5 & 5 & 5 & 5 & 5 \\

6 & 4 & 2 & 4 & 5 & 4 & 5 & 4.5 \\

6 & 5 & 2 & 4 & 4 & 4 & 4 & 4 \\

6 & 1 & 3 & 3 & 4 & 4 & 3 & 3.5 \\

6 & 6 & 3 & 5 & 4 & 4 & 5 & 4.5 \\
\bottomrule
\end{tabular}
\vspace{0.15cm}
\caption{Raw results about UX rating.}
\label{tab:ux_results}
\end{table}

Table~\ref{tab:scores} presents the accuracy scores for user responses, categorized by question. The first two columns, "Question ID" and "Paper ID," identify each question and its corresponding source paper. The subsequent columns, labeled "Score 1" through "Score 6," report the accuracy scores, providing a quantitative measure of the correctness of the answers.

\begin{table}[h!]
    \centering
    \label{tab:scores}
    \resizebox{\textwidth}{!}{%
    \begin{tabular}{ccccccccc}
        \toprule
        \textbf{question ID} & \textbf{paper ID} & \textbf{participant 1} & \textbf{participant 2} & \textbf{participant 3} & \textbf{participant 4} & \textbf{participant 5} & \textbf{participant 6} \\
        \midrule
        1 & 1 & 3.5 & 4.0 & 4.5 & 4.0 & 3.5 & 4.0 \\
        2 & 1 & 4.0 & 3.5 & 5.0 & 3.5 & 3.0 & 4.0 \\
        3 & 1 & 3.0 & 3.5 & 4.5 & 4.0 & 4.0 & 3.5 \\
        4 & 1 & 3.5 & 3.0 & 4.5 & 4.0 & 3.5 & 4.5 \\
        5 & 1 & 3.0 & 2.5 & 4.5 & 4.0 & 3.5 & 3.5 \\
        6 & 1 & 3.5 & 3.0 & 5.0 & 4.5 & 3.0 & 4.5 \\
        7 & 2 & 1.0 & 3.0 & 4.0 & 3.5 & 4.5 & 4.0 \\
        8 & 2 & 3.0 & 3.5 & 4.5 & 4.0 & 2.0 & 3.0 \\
        9 & 2 & 4.0 & 4.5 & 4.5 & 5.0 & 4.0 & 2.0 \\
        10 & 2 & 3.0 & 4.0 & 4.5 & 4.0 & 4.5 & 1.0 \\
        11 & 2 & 4.0 & 3.5 & 4.0 & 4.5 & 4.0 & 2.0 \\
        12 & 2 & 2.0 & 3.5 & 4.0 & 5.0 & 4.5 & 1.0 \\
        13 & 3 & 3.5 & 4.0 & 4.5 & 4.0 & 4.5 & 3.5 \\
        14 & 3 & 3.0 & 4.0 & 4.5 & 5.0 & 2.0 & 3.0 \\
        15 & 3 & 4.0 & 4.5 & 3.5 & 5.0 & 2.0 & 3.0 \\
        16 & 3 & 4.0 & 3.5 & 4.5 & 5.0 &  & 2.0 \\
        17 & 3 & 4.0 & 3.5 & 4.5 & 5.0 & 2.0 & 3.0 \\
        18 & 3 & 4.0 & 4.5 & 3.5 & 5.0 & 1.0 & 2.0 \\
        19 & 4 & 4.0 & 4.5 & 4.5 & 2.5 & 3.0 & 2.0 \\
        20 & 4 & 4.5 & 4.0 & 4.5 & 4.0 & 2.0 & 3.5 \\
        21 & 4 & 4.0 & 4.5 & 3.5 & 2.0 & 1.0 & 4.0 \\
        22 & 4 & 4.0 & 3.5 & 4.0 & 3.0 & 2.0 & 3.5 \\
        23 & 5 & 4.0 & 4.5 & 4.0 & 4.5 & 3.0 & 4.0 \\
        24 & 5 & 4.0 & 4.5 & 3.0 & 4.0 & 2.0 & 3.5 \\
        25 & 5 & 4.0 & 4.5 & 3.5 & 5.0 & 3.5 & 4.5 \\
        26 & 5 & 5.0 & 4.0 & 3.0 & 3.0 &  & 2.0 \\
        27 & 5 & 4.5 & 4.0 & 4.5 & 3.5 & 4.0 & 4.0 \\
        28 & 5 & 4.0 & 3.5 & 3.0 & 3.5 & 4.5 & 4.0 \\
        29 & 6 & 5.0 & 4.0 & 4.5 & 3.5 & 4.5 & 4.0 \\
        30 & 6 & 4.5 & 4.0 & 4.5 & 4.0 & 4.0 & 5.0 \\
        31 & 6 & 5.0 & 3.0 & 4.5 & 3.5 & 3.0 & 4.5 \\
        32 & 6 & 5.0 & 4.5 & 4.5 & 3.5 & 2.0 & 5.0 \\
        33 & 6 & 5.0 & 3.5 & 4.0 & 4.5 & 3.0 & 4.5 \\
        34 & 6 & 5.0 &  &  & 3.5 & 4.0 & 4.5 \\
        \bottomrule
    \end{tabular}
    }
    \vspace{0.15cm}
    \caption{Raw result of Answers Quality.}
\end{table}

\newpage
\subsubsection{Aggregated Summary Statistics}

\begin{table}[htb]
\centering
\label{table-appendix-summary}
\small
\begin{tabular}{lcccccc}
\toprule
 & \multicolumn{3}{c}{Answer Quality} & \multicolumn{3}{c}{User Experience} \\
\cmidrule(lr){2-4} \cmidrule(lr){5-7}
Metric & Estimate & $t$-value & $p$-value & Estimate & $t$-value & $p$-value \\
\midrule
\textbf{Fixed Effects} & & & & & & \\
\quad Intercept (Lv1/Lv2) & 3.106 & 15.60 & $5.25 \times 10^{-8}$ & 3.750 & 11.96 & $1.19 \times 10^{-5}$ \\
\quad Lv2 vs Reference & 0.711 & 6.07 & $6.97 \times 10^{-9}$ & -- & -- & -- \\
\quad Lv3 vs Reference & 1.139 & 9.76 & $<2 \times 10^{-16}$ & 0.375 & 1.70 & .108 \\
\midrule
\textbf{ANOVA} & & & & & & \\
\quad $F$-value & 48.44 & & $<2.2 \times 10^{-16}$ & 2.89 & & .1075 \\
\quad DenDF & 187.0 & & & 17.0 & & \\
\midrule
\textbf{Random Effects (Variance)} & & & & & & \\
\quad Subject ID & 0.1485 & & & 0.4441 & & \\
\quad Paper ID & 0.0475 & & & 0.0000 & & \\
\quad Residual & 0.4509 & & & 0.2923 & & \\
\bottomrule
\end{tabular}
\vspace{0.15cm}
\caption{Summary of mixed-effects model results.}
\end{table}

\begin{table}[htb]
\centering
\label{table-appendix-details}
\small
\begin{tabular}{lcccc}
\toprule
 & \multicolumn{2}{c}{Answer Quality} & \multicolumn{2}{c}{User Experience} \\
\cmidrule(lr){2-3} \cmidrule(lr){4-5}
Comparison & Estimate & $p$-value & Dimension & $p$-value \\
\midrule
\textbf{Pairwise Comparisons} & & & \textbf{Dimension Analysis} & \\
\quad Lv1 - Lv2 & -0.711 & $<.0001$ & Relevance & .2053 \\
\quad Lv1 - Lv3 & -1.139 & $<.0001$ & Accuracy & .7577 \\
\quad Lv2 - Lv3 & -0.428 & .0009 & Helpfulness & .0028 \\
 & & & Clarity & .5984 \\
\midrule
\textbf{Residuals Distribution} & & & \textbf{Descriptive Stats} & \\
\quad Minimum & -3.094 & & Lv2 Mean & 3.75 \\
\quad 1Q & -0.615 & & Lv3 Mean & 4.12 \\
\quad Median & 0.139 & & Improvement & +0.37 \\
\quad 3Q & 0.623 & & & \\
\quad Maximum & 3.040 & & & \\
\bottomrule
\end{tabular}
\vspace{0.15cm}
\caption{Post-hoc comparisons and descriptive statistics.}
\end{table}

\begin{minipage}{\textwidth}
\footnotesize
\textbf{Model Specifications:} \\
Answer Quality: REML criterion at convergence = 434.7, $N$ = 200 \\
User Experience: REML criterion at convergence = 50.1, $N$ = 24 \\
\textbf{Note:} \\
Reference levels: Lv1 (Without AI) for Answer Quality; Lv2 (Generic AI) for User Experience. \\
Significance codes: $^{***}p<.001$; User experience model showed singular fit (Paper ID variance = 0).
\end{minipage}

\section{Discussion}\label{disc}

\textbf{Limitations.} \texttt{QuantMind} faces several constraints: (i) potential retrieval drift under temporal or domain shifts; (ii) a limited sample size in the user study; and (iii) incomplete coverage beyond equities, with credit, rates, and derivatives requiring broader taxonomies and schemas.

\textbf{Ethics.} All sources are restricted to publicly available documents with licensing/ToS respected. User-study participants provided informed consent. Only minimal telemetry (timing and UX ratings) was logged, and all reported results are aggregated.  


\end{document}